\documentstyle[12pt,aaspp4]{article}

\def\kms{km s${}^{-1}$}
\def\ab{$\sim$}

\def\p{$\pm$}

\def\solmass{M$_\odot$}
\def\HI {H\kern0.1em{\sc i}}

\lefthead{put in names for page header}
\righthead{put in abbreviated title for page header}
\slugcomment{\it accepted for publication in the Astrophysical Journal, 1999}

\begin{document}
\null
\vspace{1cm}
\title{Obscuration of the Parsec Scale Jets in the \\ Compact Symmetric Object 1946+708}
\author{A. B. Peck\altaffilmark{1,2}, G. B. Taylor\altaffilmark{1},
J. E. Conway\altaffilmark{3}}
\altaffiltext{1} {National Radio Astronomy Observatory, P.O. Box O,
Socorro, NM 87801;\\apeck@nrao.edu,gtaylor@nrao.edu}
\altaffiltext{2}{Physics Dept., New Mexico Institute of Mining and
Technology, Socorro, NM 87801}
\altaffiltext{3}{Onsala Space Observatory, Onsala, S-439992, Sweden} 
\setcounter{footnote}{0}

\begin{abstract}
We present results of VLA and VLBA observations of the 1.420 GHz
neutral hydrogen absorption associated with the Compact Symmetric
Object 1946+708 ($z$=0.101).  We find significant structure in the gas
on parsec scales.  The peak column density in the \HI~ ($N_{\rm
HI}\simeq$2.2$\times$10$^{23}$~cm$^{-2}$(${T_s}\over{\rm 8000K}$) )
occurs toward the center of activity of the source, as does the
highest velocity dispersion ($\Delta V_{\rm FWHM}\simeq$350 \kms).  In
addition, we find that the continuum spectra of the various radio
components associated with these jets strongly indicate free-free
absorption.  This effect is particularly pronounced toward the core
and inner components of the receding jet, suggesting the presence of a
screen local to the source, perhaps part of an obscuring torus.

\end{abstract}
\keywords{galaxies:active -- galaxies:individual (1946+708) -- radio
continuum:galaxies -- radio lines:galaxies}

\clearpage
\section{Introduction}
 
The radio source 1946+708 is a Compact Symmetric Object (CSO)
associated with a m$_{\it v}$=18 galaxy at a redshift of {\it
z}=0.101\p0.001 (Stickel and K\"{u}hr 1993).  Assuming H$_0$=75
\kms~Mpc$^{-1}$, 1 mas corresponds to 1.65 pc.

CSOs are a recently identified class of radio source which are less
than 1 kpc in size, and are thought to be very young objects
($\le$10$^4$~yr, Readhead et al 1996a;Owsianik \& Conway 1998).  These
sources are uniquely well suited to investigations into the physics of
the central engines, providing in particular a means to test the
``unified scheme'' of active galactic nuclei (AGN).  The unified
scheme requires an obscuring region of atomic or molecular gas
surrounding the central engine which effectively shields the inner few
parsecs of the source from view if the radio axis lies at a large
angle to the line of sight (Antonucci 1993).

In compact sources as young as CSOs, it is reasonable to assume that
this circumnuclear material is accreting onto, and ``feeding'', the
central engine, and that this process will lead to their eventual
evolution into much larger FR II sources (Readhead et al 1996b; Fanti
et al 1995).  This infalling material is likely to form an accretion
disk (Shakura \& Sunyaev 1973).  The accretion disk will probably be
warped to some degree due to radiation pressure (Pringle 1996), and
should be predominantly molecular within a range of radii determined
by the midplane pressure in the disk (Neufeld \& Maloney 1995). This
pressure is determined by the X-ray luminosity of the central engine
and the shielding column density of hydrogen.  In regions where the
pressure falls below the critical value, both outside the radius of
the molecular region and above and below the plane of the disk, the
gas will be largely atomic.  Thus it is possible that the ``thin
disk'' called for in the accretion model of Shakura \& Sunyaev, and
the ``obscuring torus'' required for unified schemes are parts of the
same continuous toroidal structure.  We refer to the entire structure
as a ``torus,'' although it is not expected to have well defined
boundaries.  This torus would obscure the central parsecs of
sources which happen to be oriented at large angles to the line of
sight, resulting in the observed differences between the Type I
(quasar, Seyfert 1) and Type II (radio galaxy, Seyfert 2) objects.

CSOs exhibit milliarcsecond scale jets marked by steep spectrum
hotspots (Taylor et al 1996).  These jets are oriented at large angles
to the line of sight, resulting in very little Doppler boosting of the
approaching jet (Wilkinson et al 1994).  This means that the receding
jet can contain up to half of the radio continuum flux density, and
thus an obscuring torus should be detectable toward one or both
hotspots, and possibly the core (Conway 1996).  There are several
methods by which it might be possible to detect such a torus.  The
following are discussed in this paper: Atomic gas (\HI~ - $\S$3.2),
molecular gas (OH - $\S$3.3) and free-free absorption ($\S$3.4).
Interpretation of the results follows in $\S$4.

\section{Observations and Analysis}

The VLA observations were made in C Array on 6 November, 1994.  Both
the 1.420 GHz \HI~ and 1.667 GHz OH transitions were observed for a
total of 20 minutes in 3 snapshots.  Subsequently, VLBA observations
were made of the \HI~ line on 22 March, 1995 with 2 bit sampling.  A
single IF with a bandwidth of 8 MHz was observed in 512 channels,
resulting in a frequency resolution of 15.6 kHz.  This corresponds to
a velocity resolution of 4 \kms.  The 8 VLBA antennas capable of
observing at 1.286 GHz were used.  Amplitude calibration was derived
using the measurements of antenna gain and system temperature recorded
at each antenna, and refined using the calibrator 3C84 (0316+413).
Global fringe fitting was performed using the AIPS task FRING, with a
solution interval of three minutes.  Considerable attention was paid
to the bandpass calibration to eliminate a ``slope'' which was
probably due to the inapplicability of the assumed values of the
amplitude calibration signal across the band at this non-standard
frequency.  Following the standard application of bandpass calibration
in AIPS using 3C84, removal of the slope was achieved by applying a
second bandpass correction using solutions calculated from the outer
400 line-free channels of 1946+708 itself.

Continuum subtraction for the VLBA data was done in Difmap (Shepherd,
Pearson \& Taylor 1995) using a model of CLEAN components made from
several line-free channels.  Subsequent editing and imaging of all
data was also done using Difmap.  The data were then smoothed to a
velocity resolution of 12 \kms, and the absorption spectra were
analyzed using the Groningen Image Processing System (GIPSY; van der
Hulst et al 1992) and AIPS.

\section{Results}

\subsection{Radio Continuum}
The morphology and orientation of 1946+708 is well understood as a
result of a study by Taylor \& Vermeulen (1997) in which the proper
motions of the jet components were determined.  This study used data
taken at 3 frequencies over a four year period to measure the rate of
increase in angular separation between pairs of jet components.  The
components were identified from the 8 GHz VLBA radio continuum image
from Taylor \& Vermeulen (1997) shown in Figure 1a.  The rate of
expansion between components N2 and S2 was determined to be 0.17
mas/yr or 0.74$h^{-1}$c and the resulting range of angles of
inclination to the line of sight is $\theta$=65-80$\arcdeg$, where the
northern side of 1946+708 is the approaching jet.

Figure 1b shows the 1.290 GHz VLBA continuum image.  The source
components have been labeled as identified in Figure 1a.  The core is
located halfway between the northern hotspot (NHS) and southern
hotspot (SHS), which mark the ends of the approaching and receding
jets, respectively.  The observations from which the 1.290 GHz image was
made suffered a lack of short baselines owing to the inability of the
VLBA antennas at KP and FD to observe at this non-standard frequency.
Despite this, the total flux densities measured in the VLBA and VLA
observations differ by less than 2\%, indicating that there is very
little radio emission on scales of 100 to 1000 mas.

\subsection{\HI~ Absorption}

Observations were made at 1.290 GHz, the frequency of the redshifted \HI~
line, with both the VLA and the VLBA.  The resulting integrated
spectra are shown in Figure 2.  The absorption components shown are
within the errors of the redshift of the source. The optical depth of
the component centered on 1.2913 GHz is very similar in both
observations, but the lower frequency component appears to have
decreased from $\tau\simeq$ 0.050\p.007 in the VLA observation to
0.035\p.007 in the VLBA observation.  This difference could be due to
1) an actual change in opacity along the line of sight, 2) the small
amount of flux missing from the VLBA observation being absorbed on
large scales, or 3) simply to the noise in the data.  Given the known
expansion velocity of the jets, (0.03 mas in 0.37 yr) it is unlikely
that the line of sight to the background continuum has changed
sufficiently to cause this difference.

Figure 3 shows the \HI~ absorption profile toward each of 5 regions
across the source.  The regions correspond to the groups of radio
components indicated in Figure 1b.  Although there is clear evidence
of absorption toward each region, it is unclear how many distinct
components are present in each profile.  With the exception of the
profile toward the northern hot spot (NHS+N1), a single Gaussian
function has been fitted to each profile.  The optical depths ($\tau$)
determined by these fits are presented in Table 1.  The systemic velocity
obtained from optical observations of both emission and stellar
absorption lines (Stickel \& K\"{u}hr 1993) is 30279\p300 \kms.  Thus
all of the \HI~ absorption features reported here are within one sigma
of the systemic velocity.

In addition to fitting Gaussian functions to the integrated \HI~
absorption profiles in the boxes shown, fits have also been made at
each pixel across the source where the continuum level is higher than
50 mJy.  In the regions roughly corresponding to the NHS, N1, N2, and
N3 components, two Gaussians were fitted as shown in Figure 4.  These
functions were limited to a maximum FWHM of 160 \kms~ centered within
100 \kms~ of the velocities shown for NHSa and NHSb in Table 1.  The
first row of plots in Figure 4 contain the parameters obtained from
fits to the 29970 \kms~ line.  Pixels with a peak signal to noise
ratio less than 2 have been blanked.  The absorption occurs across all
of the continuum source, with velocity centroids ranging from 29950 to
30070 \kms ~(Figure 4a).  The linewidth distribution shown in Figure
4b indicates a dramatic increase in velocity dispersion toward the
inner receding jet (S2+S3), where the FWHM is $\sim$150 \kms.  The
optical depth is fairly uniform ($\tau\sim0.06$, Figure 4c) with the
exception of the steep gradient toward the receding jet where the
optical depth rises to 0.18 on the eastern edge.  This corresponds to
a region of extremely narrow FWHM, and so might be attributable to bad
fits to the data.

Due to the relative weakness of the line centered on 30900 \kms, fits
with a peak signal to noise ratio $\ge$ 1.5 have been allowed.  These fits
appear in the second row in Figure 4.  The central velocities are
fairly uniform (V\ab30070 \kms) across the portions of 1946+708 toward
which this line is present, as are the FWHM ($\Delta V \sim 50$ \kms)
and optical depth ($\tau\sim0.04$).

\subsection{OH Abundance}

Figure 5 shows the VLA spectrum centered on 1.514 GHz, the frequency of
the redshifted hydroxyl line.  The relationship of the optical depth to the
abundance of OH along the line of sight is
$$N(OH)\simeq2.27\times10^{14}T_{ex}\tau\Delta V$$ (O'Dea \& Baum
1987).  The excitation temperature ($T_{ex}$) of OH in galactic clouds
tends to be much less than the kinetic temperature, in most cases
between 4 and 8 K (Dickey, Crovisier \& Kaz\'es 1981;Turner 1985), so
a value of 8 K is adopted.  We assume a $\Delta$V of 100 \kms~ based
on the average \HI~ linewidths.  For the upper limit on optical depth of
$\tau\sim$0.006 shown, this results in an upper limit on the column
density of 1.089$\times$10$^{15}$$\left(T_{ex}\over8{\rm
K}\right)\left(\Delta V\over{100 {\rm km/s}}\right)$~cm$^{-2}$.

Although this is not a particularly stringent upper limit, it is
consistent with the non-detections of OH in similar sources (i.e. Cygnus
A, Conway 1998).  A number of theories have been proposed to explain
the apparent lack of molecular absorption (Barvainis \& Antonucci
1994; Black 1998), but the simplest is that the extended circumnuclear
structure referred to as a torus is predominantly atomic (Maloney
1998).

\subsection{Radio Continuum Spectra} 

By tapering a 5 GHz image of 1946+708 taken using the VLBA on 1995
Sept. 3 (Taylor \& Vermeulen 1997) we obtained an image with
resolution matched to our 1.290 GHz continuum image.  These two images
were then combined to generate an image of the spectral index
distribution across the source (Fig.~6).  At either end of the source
a steep spectrum is found, reaching $\alpha = -0.99 \pm 0.1$ (where
$S_\nu \propto \nu^\alpha$) in the SW hot spot.  In the NE hot spot
the spectrum is flatter ($\alpha = -0.29$), and at the center of the
source and extending across the inner southern jet, the spectral index
is quite flat ($\alpha = 0$).  These substantial deviations from the
expected spectral index of $-$1 for optically thin synchrotron
emission indicate that the radio emission is absorbed at frequencies
below $\sim$5 GHz.

The overall spectrum of the nucleus of 1946+708 from single dish
observations (Taylor \& Vermeulen 1997 and references therein) from 330
MHz to 100 GHz is shown in Fig.~7.  From the VLBA observations (this
paper and Taylor \& Vermeulen 1997) we can study the spectra of groups
of components between 1.3 and 15 GHz, though the dramatic change in
resolution and ($u,v$) spacings over this range of frequencies can
result in the flux density of some larger components (especially SHS)
being underestimated at 8 and 15 GHz.  We see that the spectra of the
various components differ substantially.  Most noteworthy is the inner
portion of the receding jet.  The spectrum of this group of components
(S2 + S3) is fairly steep between 8 and 5 GHz, but flattens
dramatically between 5 and 1.3 GHz.

\section{Discussion}

A number of sources have been discovered which exhibit clear evidence
of one or more of the features associated with an obscuring torus.  We
list a few recent examples.  Herrnstein et al (1997) have determined
the size and shape of the warped molecular disk in NGC 4258 as traced
by maser spots.  In this source, the masing disk extends from 0.13 pc
to 0.25 pc from the central engine.  Evidence of a circumnuclear torus
of atomic gas has been seen in Cygnus A and Hydra A.  In Cygnus A, \HI~
absorption measurements with the VLBA indicate a torus with a radius
of \ab 50 pc (Conway 1998).  In Hydra A, evidence is found for
free-free absorption toward the core and inner jets, as well as an \HI~
torus with a scale height of $\sim$30 pc (Taylor 1996).

Additional sources show evidence of a toroidal structure, but the
distribution of maser spots or \HI~ absorption is not as well
understood.  For example, megamasers are also seen in the Circinus
galaxy (Greenhill et al 1997), NGC 3079 (Baan \& Haschick 1996) and
NGC 4945 (Greenhill et al 1997).  At least one of these, NGC 3079,
also exhibits \HI~ absorption toward the core (Satoh et al 1997).
Observations of NGC 1068 suggest evidence of three phenomena,
consisting of OH and H$_2$O masers, \HI~ absorption, and a compact
source of free-free emission (Gallimore, Baum \& O'Dea 1996).  In this source,
the maser emission indicates that the molecular disk makes a large
angle with the radio axis (Greenhill \& Gwinn 1997) and so might be
more strongly warped than that of NGC 4258.

Figure 8a shows a cartoon of what the circumnuclear structure could
look like.  Some notable simplifications have been made.  For example,
it is not necessary for the toroidal structure to be perpendicular to
the jet axis, the ``clumps'' of denser gas are unlikely to be uniform
in size, and the degree of warp in the disk can vary greatly (Maloney
1998).  Also, the relative scale heights and radii of the various
regions are not very well constrained.  This model represents only the
central $\le$100 pc of the source, and does not address the
possibility of an extended disk or torus associated with the host
galaxy, although this could conceivably be part of the same continuous
structure.  The extent of the molecular disk is likely to be on the
order of a parsec, and the region of atomic gas beyond that probably
extends some 50-100 pc from the central engine.  This outer region,
although predominantly neutral atomic gas, should contain a
significant fraction of free electrons (1 - 10\%, Maloney, Hollenbach
\& Tielens 1996), and also probably contains clumps or clouds of denser
gas which could harbor molecular gas.

Applying this model to 1946+708, the atomic region in the torus
appears to be at least 50 pc in radius and 20 pc in scale height.  The
free-free absorption is probably due to ionized gas in this torus, and
to the central ionized region along the line of sight to the southern,
receding jet.  Our line of sight to the jet components is shown in
Figure 8b.  

\subsection{The ionized gas}

Three of the regions in 1946+708 exhibit a flattening of the continuum
spectrum between 1.3 and 5 GHz (Fig. 7); NHS+N1, C+N5+S5, and S2+S3.
There are generally two mechanisms invoked to produce such a low
frequency turnover in compact extragalactic radio sources: (1)
synchrotron self-absorption and (2) free-free absorption by a screen
of ionized gas.  Under equipartition conditions and for uniformly
filled components, synchrotron self-absorption becomes important at
GHz frequencies for components with brightness temperatures ($T_b$)
approaching the mean kinetic energy ($10^{10}$ K) of the radiating
relativistic electrons (Williams 1963).  Thus synchrotron
self-absorption is only likely for the northern hot spot (NHS+N1),
where $T_b = 4 \times 10^8$ K, and the unresolved core ($T_b > 10^9$
K) in 1946+708.  Since the core makes only a minor contribution ($<$
10 \%) to the flux density in the inner parts of the source at 1.3 and
5 GHz, however, it seems likely that free-free absorption is the
dominant mechanism for the low frequency turnover in both C+N5+S5 and
S2+S3.  For a pure H plasma the free-free optical depth is given by:
$$
\tau_{\rm ff}  \simeq 5 \times 10^{-8}\,T_4^{-3/2}\,n_{\rm
  e}^2\,\nu_9^{-2}\,g\,d_{\rm pc},
$$(
Scheuer 1960;Mezger \& Henderson 1967; Levinson, Laor, \& Vermeulen 1995)
where $d_{\rm pc}$ is the path length through the absorbing medium in
parsecs, $T_4$ is the gas temperature in units of 10$^4$ K, $n_{\rm
e}$ is the electron density in cm$^{-3}$, $\nu_9$ is the frequency in
GHz, and $g$ is the thermal average Gaunt factor. This factor is of
order 5 for $T_4$ = 1 and $\nu_9$ = 1.3 (Spitzer 1978).  We estimate
the expected flux densities at 1.3 GHz for the two free-free absorbed
regions based on the observed spectral indices between 5 and 15 GHz.
These values are shown in Table 2.  The free-free optical depth,
column densities, and electron densities are also calculated, assuming
a path length of 50 pc through the obscuring material toward C+N5+S5
and S2+S3, and a temperature $T_4$ = 0.8.  If the ionized gas is
associated with the \HI~ torus, as the distribution of spectral
indices in Figure 6 and the high column density toward the inner
receding jet suggest, the expected ionized fraction is 1-10\% (Maloney
et al 1996).  The column densities shown in Tables 1 and 2 indicate an
ionization fraction around 11\% toward the core and 30\% toward the
inner receding jet.  Although these large percentages could be due to
incorrect assumptions in the calculation of column densities, this
difference could indicate that along the line of sight to S2+S3, the
inner receding jet, we are also probing the region between the inner
torus and the jet, where the ionization fraction should be much
higher.

\subsection{The atomic gas}
In all profiles except (NHS+N1) shown in Figure 3, there appear to be
two or more lines which are very close in velocity or are blended.
The top row in Figure 4 shows that single Gaussian functions centered
on 30040\p70 can be fit to the profiles in almost all regions across
the source.  This uniformity does not indicate, however, that a single
\HI~ feature is present in front of the entire radio continuum source,
but rather that when numerous components are present, their velocities
straddle this value.  This can be seen in the range of $\Delta$V
(30-150 \kms).  If the velocity dispersion seen in the profiles is
indicative of rotation, this central velocity of 30040\p70 \kms~ is likely
to be the true systemic velocity.

We can compare the \HI~ data from 1946+708 to sources in which we know
that the \HI~ absorption is taking place in colder gas in a kpc scale
torus or in the host galaxy.  In these cases, we see multiple lines
along each line of sight, all of which are quite narrow and distinct
from one another.  In the case of Centaurus A where a kpc scale dust
lane intersects our line of sight to the core, at least three \HI~
absorption components are present, each with a FWHM of only 7 to 12
\kms~(van der Hulst, Grolisch \& Haschick 1983; Peck \& Taylor 1998).
Similarly, toward the center of our own Galaxy, Liszt et al (1983)
find multiple components with an average FWHM $<$20 \kms.  In
contrast, although there are indications for substructure within the
1946+708 absorption profiles (see Section 3.2), the individual
components are considerably broader than those in the galactic or
Centaurus A absorption profiles.  Furthermore these components, rather
than being distinct, are overlapping in velocity. This blending of the
absorption lines is probably due to high velocity dispersion between
clouds or clumps in a hot medium.  These profiles suggest that the HI
gas in 1946+708 lies in a considerably more energetic environment,
such as that found within a few hundred parsecs of the central engine
of an AGN.  This model is further supported by variations in the
column densities over regions on the order of 10 pc, with the highest
values centered on the core of the radio source.

The optical depth is calculated using the integrated line amplitude
and the integrated continuum flux within each box outlined in Figure
3.  The column density is calculated using
$$N_{\rm HI}\simeq1.823\times10^{18}{\rm
cm^{-2}K^{-1}(km/s})^{-1}\times(\tau/f)\times T_{s}\times\Delta V$$
assuming a covering factor ($f$) of 1, and a spin temperature (T$_s$)
of 8000 K.  Models predict that atomic gas irradiated by hard X-rays
should have a stable equilibrium temperature of $\sim$8000 K (Maloney
et al. 1996).  According to the calculations of Maloney, Hollenbach
and Tielens (1996), circumnuclear \HI~ gas becomes self shielding to
the hard X-rays which heat it for column densities of 10$^{22}$ cm$^{-2}$~ and
above. We can therefore ask whether the kinetic temperature of 8000 K
can be maintained through such a gas column, and hence whether
solutions with this temperature and column density are
self-consistent. Clearly the result depends entirely on the (presently
unknown) hard X-ray luminosity of 1946+708. However, we can say that
if this X-ray luminosity is comparable to that of Cygnus A (which also
has a comparable inferred \HI~ column), then for gas which lies within
1 kpc of the central engine, the X-ray heating would be sufficent to
keep all of the \HI~ column at 8000 K (Blanco \& Conway 1996, Conway
\& Blanco 1995).  Although there might also be cooler gas present at
the interface between the molecular gas and \HI~ in the dense clumps
within the torus, the lack of molecular absorption lines suggests that
the covering factor of these dense clouds is quite small, and
therefore the contribution to the overall \HI~ absorption from these
regions would be negligible.

Given an assumed distance of 100 pc from the central engine, the
inferred \HI~ column density (assuming T$_s$ = 8000 K and a maximum
atomic gas filling factor of unity) implies a neutral hydrogen density
of $n_{\rm HI}\sim$8$\times$10$^2$ cm$^{-3}$.  At this distance, given
the source radio flux density, the increase in spin temperature due to
radiative excitation should be less than 15\% (Bahcall \& Ekers 1969).
Hence the assumption that the spin temperature is about equal to the
kinetic temperature is self consistent in our modelling and we will
assume this in our calculations.

The regions of peak absorption intensity toward the northern hotspot
are separated by \ab2 mas (33$h^{-1}$~pc).  It seems unlikely that
this could be due to the rotation of gas in a circumnuclear torus,
because it would indicate that the torus is offset by 30$\arcdeg$ from
the radio axis, rather than the expected 90$\arcdeg$, and that the
inner radius of the torus has a velocity of only 65 \kms.  The offset
of the peaks of these two lines is more likely to be due to the
presence of ``clumps'' in the toroidal structure surrounding the
central engine.  These narrow lines could be dense compact regions of
\HI~ in the host galaxy, and not associated with the AGN, though they
are still substantially broader than typical galactic \HI~ absorption
lines.

The high column density and velocity dispersion of the gas toward the
core is more indicative of a rotating torus.  The signal to noise
ratio in the present observations was not high enough to determine the
spatial offset between the outermost velocity components, and so the
simplest model of a torus perpendicular to the radio axis is assumed.
Given that the high opacity region covering the core is less than 50
pc in height, and that the broad absorption covers much of the
counterjet which is at 65-80$\arcdeg$~to the line of sight, the
obscuring torus is probably at an angle of order 20$\arcdeg$~ from
edge-on with a radius of at least 50 pc (see Figure 8b).  Assuming
gravitationally bound gas, possibly in rotation, and a relatively
thick disk able to cover both the counterjet and the core, then the
dispersion velocity along the line of sight must be comparable to the
rotational velocity or total random velocity within the disk.  The 350
\kms~FWHM of the gas at 25$-$50 pc then implies, from rough virial
arguments, a central mass of M$\sim$5$\times$10$^8$ \solmass.  Without
a detailed model, this estimate is highly uncertain, but it is
consistent with the central masses inferred in active galactic nuclei
by other methods.

\section{Conclusion}

Neutral hydrogen is present toward all of the radio components in the
Compact Symmetric Object 1946+708.  Narrow lines seen toward the
northern hotspot which marks the approaching jet are probably due to
small clouds of \HI~ associated with an extended ``clumpy'' torus of
warm gas.  The high velocity dispersion and column density toward the
core of the source, however, are indicative of fast moving
circumnuclear gas, perhaps in a rotating toroidal structure.  Further
evidence for this region of high kinetic energy and column density is
found in the spectra of the jet components, which indicate a region of
free-free absorption along the line of sight toward the core and inner
receding jet.  The most likely scenario to explain these phenomena
consists of an ionized region around the central engine, surrounded by
an accretion disk or torus with a radius of at least 50 pc which is
comprised primarily of atomic gas.

\begin{acknowledgements}

The authors thank the referee, Jack Gallimore, and Miller Goss for
helpful suggestions and discussion.  The National Radio Astronomy
Observatory is a facility of the National Science Foundation operated
under a cooperative agreement by Associated Universities, Inc.  AP is
grateful for support from NRAO through the pre-doctoral fellowship
program.  This research has made use of the NASA/IPAC Extragalactic
Database (NED) which is operated by the Jet Propulsion Laboratory,
California Institute of Technology, under contract with the National
Aeronautics and Space Administration.

\end{acknowledgements}

\null\clearpage


\begin{figure}
\vspace{16cm}
\includegraphics{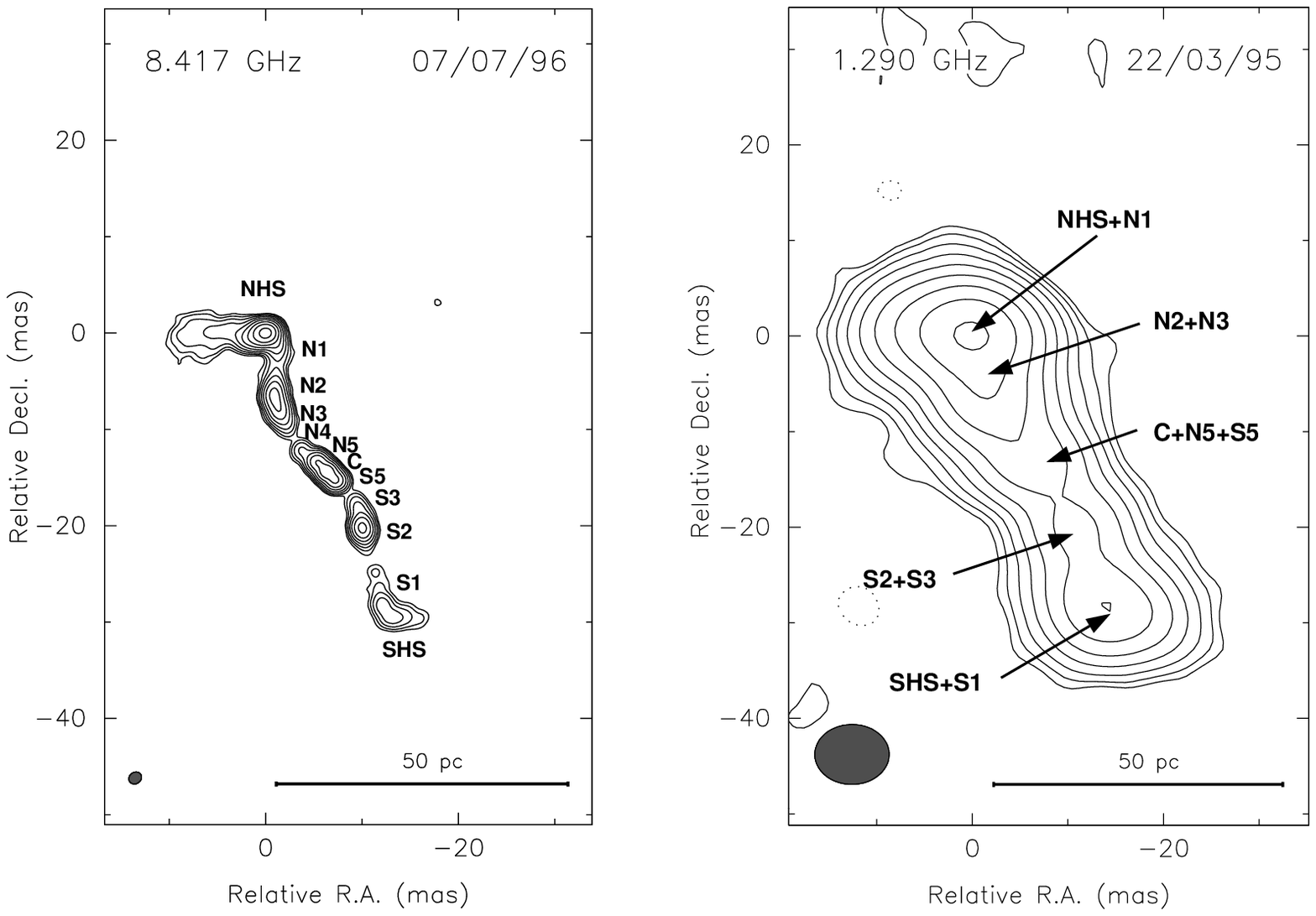}
\figcaption{{\bf}1$a$.  Continuum image of 1946+708 at 8.42 GHz.  Contours
are drawn logarithmically at factor 2 intervals from 0.25
mJy/beam. The beam size is 1.4 $\times$ 1.2 mas with position angle
$-$55$\arcdeg$.  10 mas is equivalent to 16.5$h^{-1}$~pc.  NHS and SHS
denote the northern and southern hot spots, respectively.  The
northern side of the jet is the approaching side.  {\bf}1$b$.  Continuum
image of 1946+708 at 1.29 GHz.  Contours are shown at -2, 2, 4, 8, 16,
32, 64, 128 and 256 mJy/beam.  The beam size is 7.8 $\times$ 6.4 mas
with position angle 89$\arcdeg$.  The rms noise in the image shown is
0.7 mJy/beam.  The continuum peak is 432 mJy/beam, and the total flux
in the source is 1.014 Jy.
\label{fig1}}
\end{figure}

\begin{figure}
\vspace{12cm}
\includegraphics{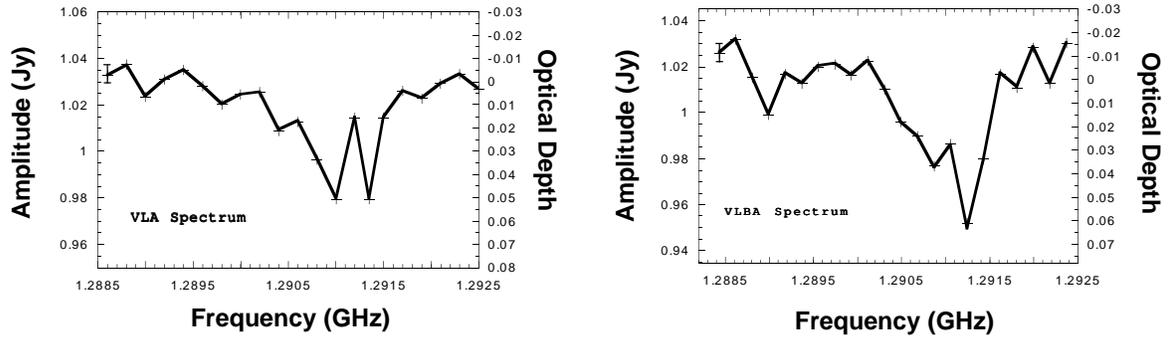}

\figcaption{VLA and VLBA \HI~ absorption spectra. The first panel shows the
VLA spectrum taken in November 1994, in C Array.  The second panel
shows the integrated spectrum over the entire source taken with the
VLBA in March 1995.  A typical error bar is shown on the first point
in each plot.
\label{fig2}}
\end{figure}

\begin{figure}
\vspace{18cm}
\includegraphics{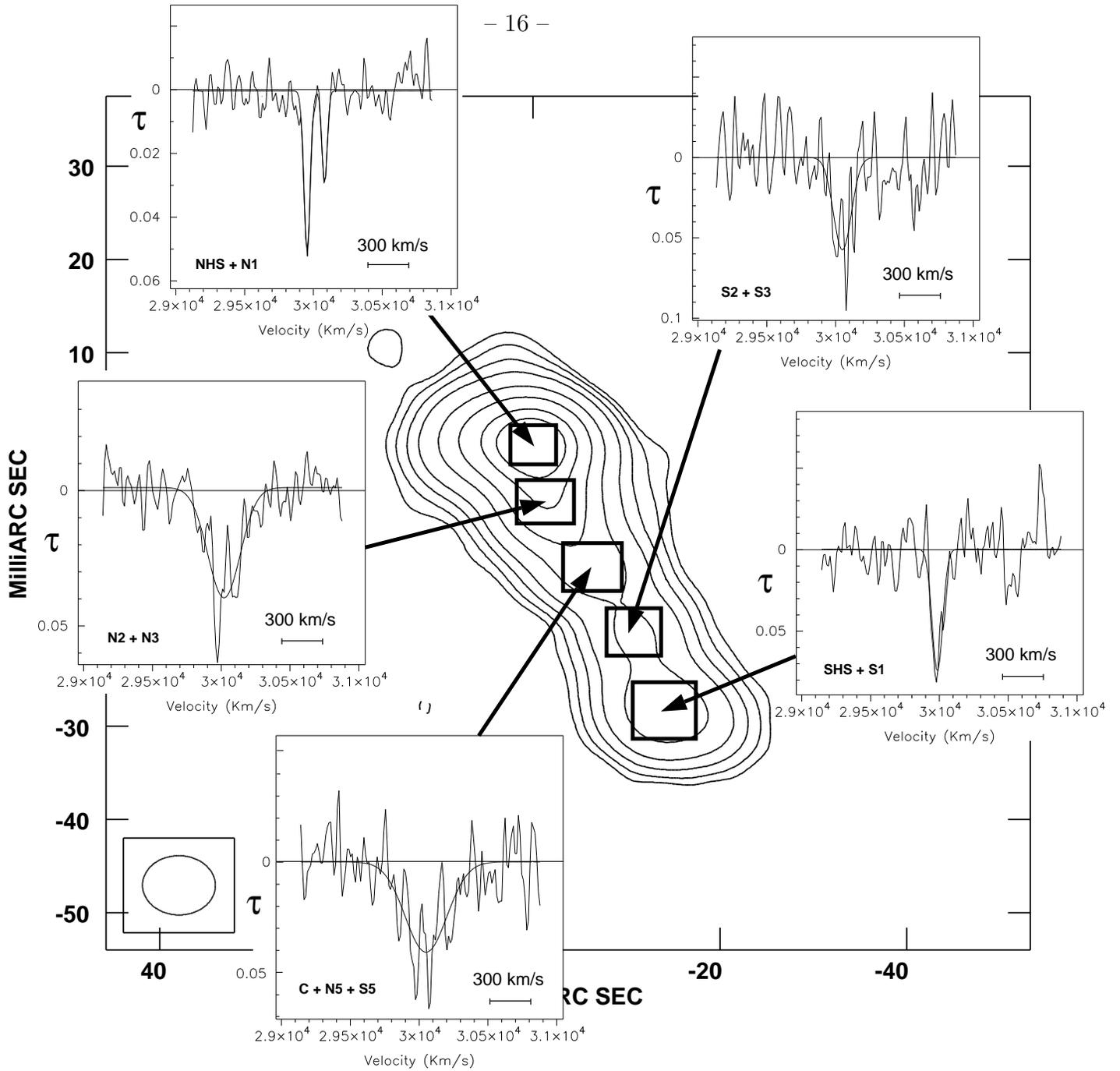}

\figcaption{ Integrated profiles of \HI~ absorption in 5 regions
corresponding to radio components identified at 8 GHz in Taylor \&
Vermeulen (1997).  The profile toward the continuum peak (NHS+N1)
exhibits two distinct lines separated by 120 \kms, while the rest
contain several blended lines, so two Gaussian functions were fitted
to the data in (NHS+N1), and a single Gaussian was fit to each of the
other profiles.  The data were averaged to 12 \kms~ resolution for the
Gaussian fits shown, and subsequently Hanning-smoothed to reduce the
noise in the plots.  The rms noise in the profiles is \ab2
mJy/beam/channel.
\label{fig3}}
\end{figure}

\begin{figure}
\vspace{18cm}
\includegraphics{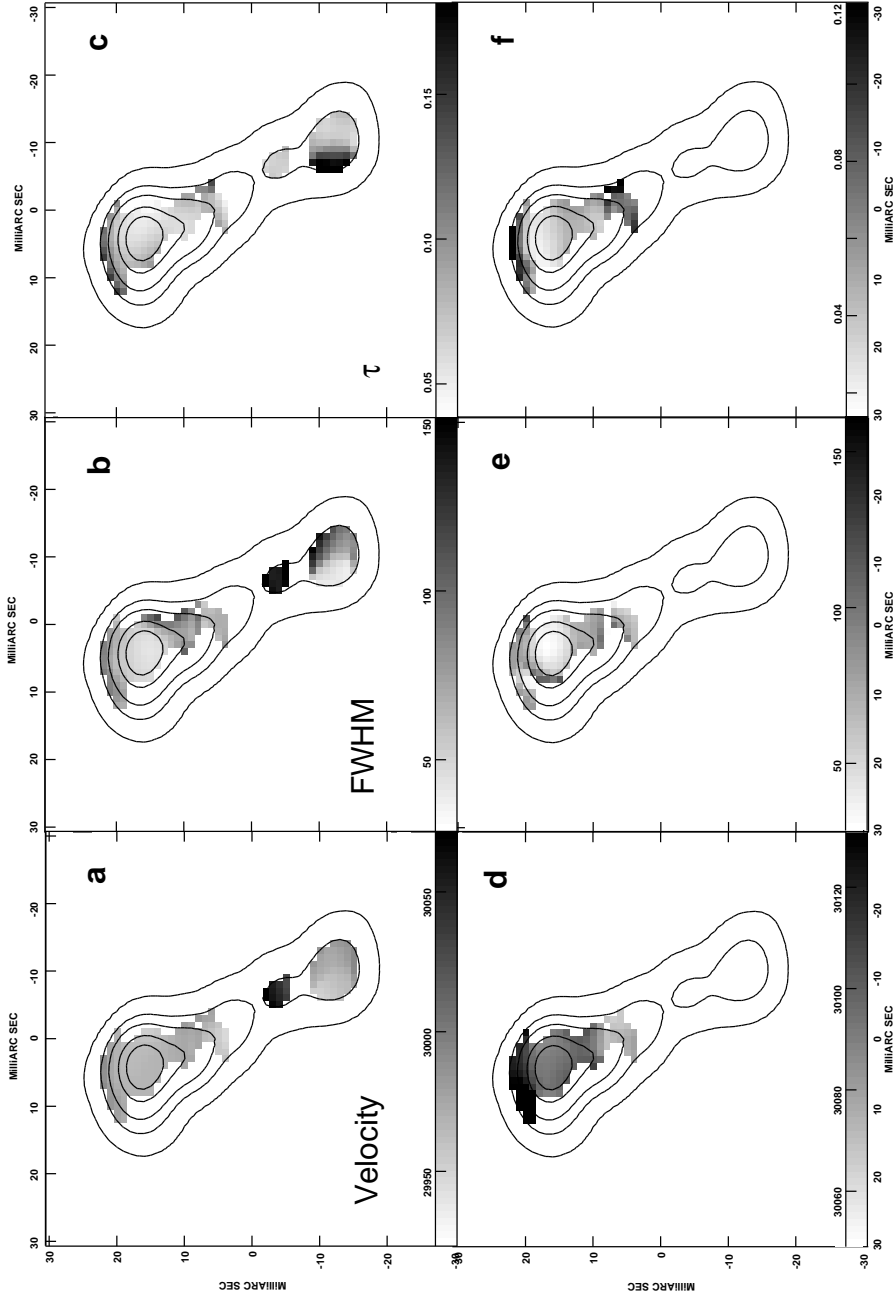}

\figcaption{The results of Gaussian functions fitted at each pixel
where the continuum emission is at least 50 mJy.  Grayscale values
increase from lighter to darker.  Panel 4a shows the velocity field of
the line centered on 29970 \kms.  Grayscales range from 29950 to 30070
\kms.  Panel 4b shows the FWHM of the line centered on 29970 \kms.
Grayscales range from 30 to 150 \kms.  Panel 4c shows the optical
depth of the 29970 \kms~ line.  Grayscales range from $\tau$=0.04 to
0.18.  Panel 4d shows the velocity field of the line centered on 30090
\kms. Grayscales range from 30070 to 30150 \kms. Panel 4e shows the
FWHM of the line centered on 30090 \kms.  Grayscales range from 20 to
100 \kms.  Panel 4f shows the optical depth of the 30090 \kms~ line.
Grayscales range from $\tau$=0.02 to 0.12.
\label{fig4}}
\end{figure}

\begin{figure}
\vspace{12cm}
\includegraphics{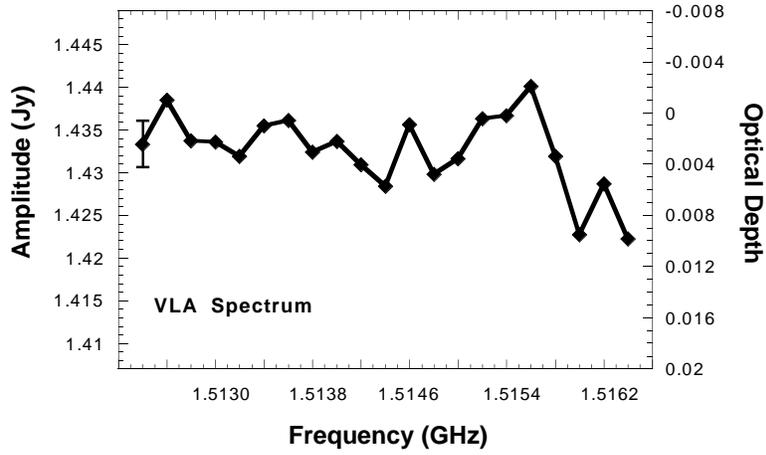}

\figcaption{VLA spectrum of 1946+708 at 1.514 GHz, the frequency of
the OH line at the redshift of {\it z}=0.101.  A typical error bar is
shown on the first point in the plot.  The profile shows an upper limit
on the OH optical depth of $\tau$=0.006, indicating an upper limit on
the OH column density of 1.089$\times$10$^{15}$ cm$^{-2}$.
\label{fig5}}
\end{figure}

\begin{figure}
\vspace{18cm}
\includegraphics{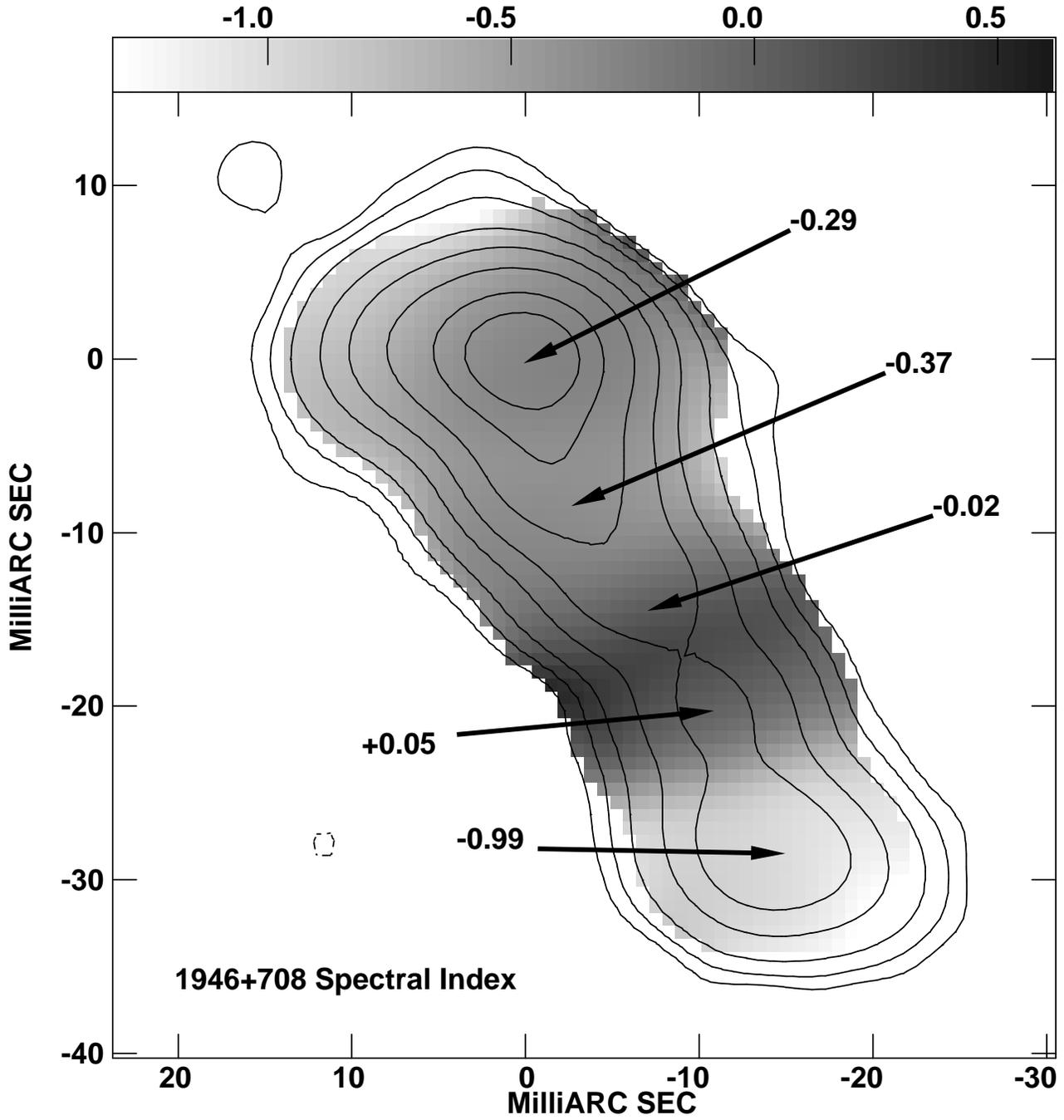}

\figcaption{Distribution of spectral indices across 1946+708 overlayed on
continuum contours, (where $S_\nu \propto \nu^\alpha$).  The spectrum
flattens substantially toward the core and inner receding jet.
\label{fig6}}
\end{figure}

\begin{figure}
\vspace{16cm}
\includegraphics{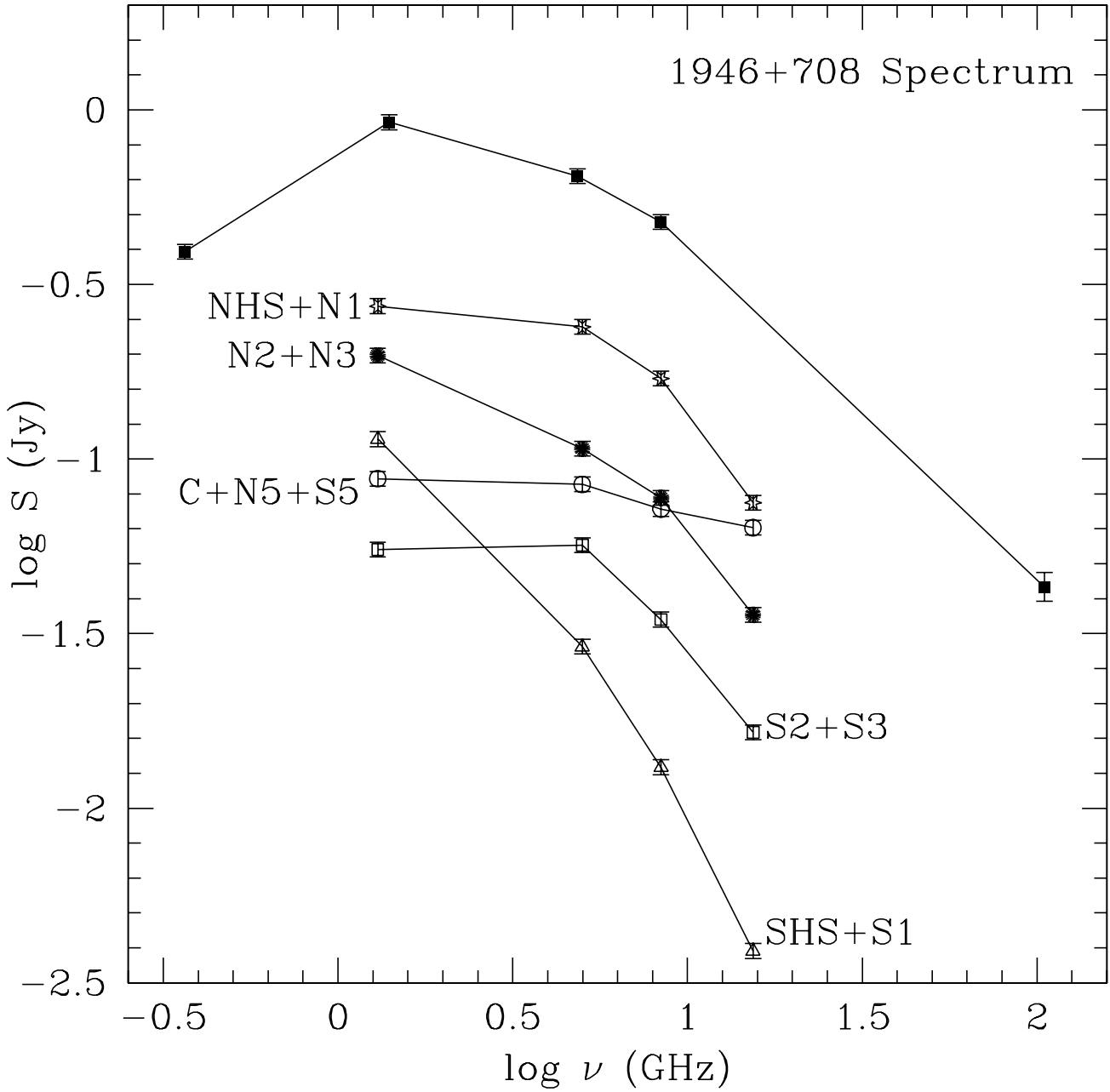}

\figcaption{Continuum spectra of the various radio components in
1946+708.  The spectra of components C+N5+S5 and S2+S3 show clear
evidence of free-free absorption, indicating the probable presence of
ionized material local to the source.
\label{fig7}}
\end{figure}

\begin{figure}
\vspace{16cm}
\includegraphics{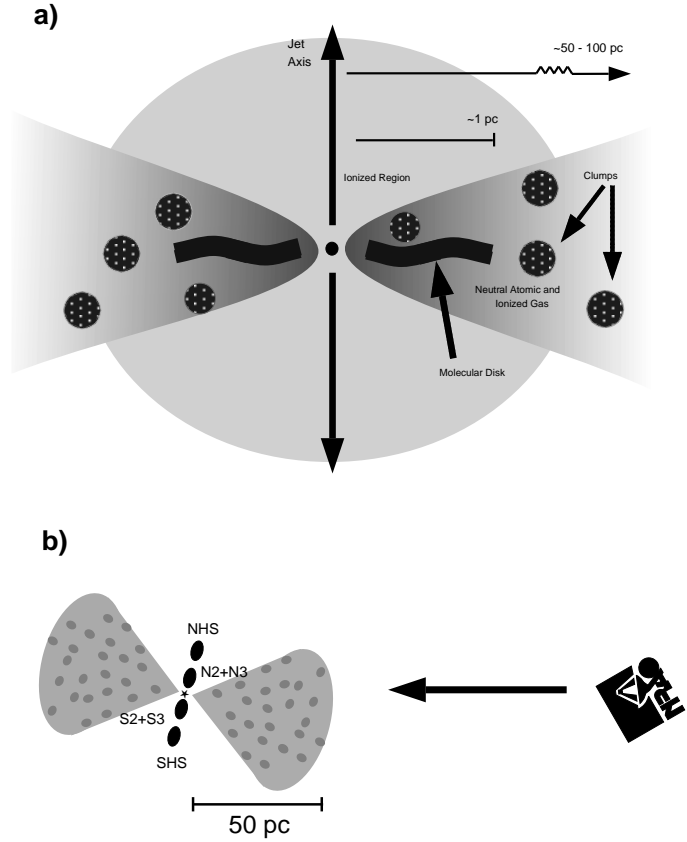}
\figcaption{a) A cartoon showing the possible environment in the central
parsecs of AGN.  Some notable simplifications have been made.  For
example, it is not necessary for the toroidal structure to be
perpendicular to the jet axis, the ``clumps'' of denser gas are
unlikely to be uniform in size, and it is believed that the degree of
warp in the disk can vary greatly.  b) The line of sight to the jet
components in 1946+708.
\label{fig8}}
\end{figure}

\clearpage

\begin{table}
\begin{center}
\begin{tabular}{crcccc}
 & Amplitude & Central Velocity & FWHM & & N$_{\rm HI}$\tablenotemark{a} \\
Components & (mJy)~ & (\kms) & (\kms) & $\tau$ & (10$^{22}$ cm$^{-2}$) \\
\hline
NHS+N1a & 13.4\p2.2 & 29970\p2.5 & 47.8\p5.9 & 0.057\p.009 &
4.1\p0.8 \\
NHS+N1b & 7.8\p2.4 & 30090\p3.9 & 40.1\p9.3 & 0.033\p.010 &
2.0\p0.7  \\
N2+N3 & 7.5\p1.2 & 30010\p12 & 273\p29 & 0.041\p.006 & 17\p3.0 \\
C+N5+S5 & 3.8\p1.0 & 30040\p23 & 357\p62 & 0.042\p.010 &
23\p6.7 \\
S2+S3 & 4.8\p1.7 & 30050\p16 & 150\p39 & 0.060\p.020 &14\p5.6  \\
SHS+S1 & 7.5\p2.1 & 29970\p6.9 & 88.3\p17 & 0.070\p.019 &
11\p2.9 \\

\hline
\end{tabular}
\tablenotetext{a}{Assuming a spin temperature of 8000 K and a covering
factor of 1.}
\end{center}
\tablenum{1}
\caption{Gaussian Functions fitted to Absorption Profiles in
Each Region}
\end{table}

\begin{table}
\begin{center}
\begin{tabular}{lccccc}
 & Measured & Estimated True & & N$_{\rm e}$\tablenotemark{a} & Density\tablenotemark{a}\\
Component & Flux Density & Flux Density & $\tau_{ff}$ & (cm$^{-2}$) & (cm$^{-3}$)\\
\hline
C+N5+S5 & 88 mJy & 126 mJy & 0.4 &  3.0$\times$10$^{22}$ & 200 \\
S2+S3 & 55 mJy & 200 mJy & 1.3 & 5.3$\times$10$^{22}$ & 350 \\

\hline
\end{tabular}
\tablenotetext{a}{Assuming a temperature of 8000 K and a
pathlength of 50 pc.}
\end{center}
\tablenum{2}
\caption{Parameters determined from free-free absorption toward the
core and receding jet.}
\end{table}
\clearpage

\end{document}